\documentclass[aps,prl,groupedaddress,twocolumn]{revtex4-2}
\usepackage[fleqn]{amsmath}
\usepackage{float}
\usepackage{graphicx}
\usepackage{graphics}
\usepackage{epstopdf}
\usepackage{amssymb}
\usepackage{amsmath}
\usepackage{bm}
\usepackage{braket}
\usepackage{color}
\usepackage[colorlinks,citecolor=blue]{hyperref}
\usepackage{changes}
\usepackage{CJK}
\usepackage{lineno}

\begin{document}
%\linenumbers
\title{Observation of interaction-induced fast Thouless pumping of solitons}
\author{Yuqing Li$^{1,3,4}$}
\thanks{These authors contribute equally to this work.}
\author{Jinxiong Jia$^{2,3}$}
\thanks{These authors contribute equally to this work.}
\author{Yunfei Wang$^{1,3}$}
%\thanks{These authors contribute equally to this work.}
\author{Huiying Du$^{1}$}
\author{Zhong An$^{5}$}
\author{Zhenhua Qiao$^{2,3}$}
\thanks{Corresponding author: {qiao@ustc.edu.cn}}
\author{Liantuan Xiao$^{1,3}$}
\author{Suotang Jia$^{1,3}$}
\author{Qian Niu$^{2}$}
%\author{*** ***$^{6}$}
\author{Jie Ma$^{1,3,4}$}
\thanks{Corresponding author: {mj@sxu.edu.cn}}
\affiliation{
$^1$State Key Laboratory of Quantum Optics Technologies and Dvices, Institute of Laser Spectroscopy, College of Physics and Electronics Engineering, Shanxi University, Taiyuan 030006, China\\
$^2$International Centre for Quantum Design of Functional Materials, and Department of Physics, University of Science and Technology of China, Hefei, Anhui 230026, China\\
$^3$Hefei National Laboratory, Hefei 230088, China\\
$^4$Collaborative Innovation Center of Extreme Optics, Shanxi University, Taiyuan 030006, China\\
$^5$College of Physics and Hebei Advanced Thin Film Laboratory, Hebei Normal University, Shijiazhuang, Hebei 050024, China}

\begin{abstract}
Thouless pumping provides a paradigmatic platform for studying the effects of interactions on topological transport in periodically driven systems. However, most studies have been constrained by adiabatic conditions, which preclude exploration of interaction-driven novel topological states at high driving frequencies. Here, we experimentally investigate the interplay between interaction and modulation frequency in Thouless pumping realized in a periodically modulated lattice in momentum space of atomic Bose-Einstein condensate. We observe fast Thouless pumping of matter-wave solitons at intermediate interactions, with no counterpart in the non- or weakly interacting regimes. Beyond the boundary of topological phase transition induced by interaction, nonadiabatic quantized pumping of solitons emerges at high modulation frequencies over a broad interaction range, in good agreement with theoretical calculations, while the solitons remain trapped in the low-frequency adiabatic pumping regime. Our work opens new avenues for accelerating topological transport in driven quantum systems and engineering fast topological devices.

\end{abstract}

\maketitle

Interactions in periodically driven systems can give rise to emergent collective phenomena, ranging from time-crystalline dynamics to correlated Floquet topological phases that have no equilibrium analog
~\cite{Xiao:RMP2010, Cooper:RMP2019, Citro:NRP2023,2019-Floquet,2020-time_crystals}.
A paradigmatic platform for exploring interaction effects in one-dimensional (1D) periodic systems is the Thouless pump, which can be regarded as a dynamical version of the integer quantum Hall effect~\cite{Thouless:PRL1982, Thouless:PRB1983,Niu:JPA1984}. In such pumps, the quantized transport per cycle of adiabatic evolution is characterized by Chern number of occupied energy bands. While initially proposed in solid-state systems~\cite{Niu:PRL1990}, Thouless pumps have been extensively realized in a wide range of physical systems in non-interacting regime, such as ultracold atoms~\cite{Nakajima:NP2016, Lohse:NP2016, Nakajima:NP2021}, photonic waveguides~\cite{Kraus:PRL2012, Cheng:NC2022, Song:SA2024}, superconducting processors~\cite{Liu:NC2025}, acoustics~\cite{Cheng:PRL2020, You:PRL2022} and mechanics~\cite{Grinberg:NC2020}.

Recent experiments have made significant progress in exploring how interparticle interactions modify the behavior of Thouless pumping. In photonic systems, the quantized integer and fractional Thouless pumping of solitons have been demonstrated under focusing Kerr nonlinearity~\cite{Jurgensen:Nat2021, Jurgensen:NP2023}, stimulating more ongoing theoretical interests~\cite{Jurgensen:PRL2022, Mostaan:NC2022, Fu:PRL12022, Fu:PRL22022, Ravets:PRL2025, Xu:PRL2025, Jurgensen:PRL2025, Wu:Aix2025}. Strong Hubbard interaction-induced breakdown and emergence of Thouless pumping have been successively observed in ultracold atoms~\cite{Walter:NP2023, Viebahn:PRX2024}.
However, implementations of quantized Thouless pumping both with and without interactions have been largely restricted by adiabatic conditions, which require the slow cyclic evolution of system's parameters and hinder practical applications~\cite{Privitera:PRL2018, Song:SA2024}.

\begin{figure} [!hbt]
\includegraphics[width=8.5cm]{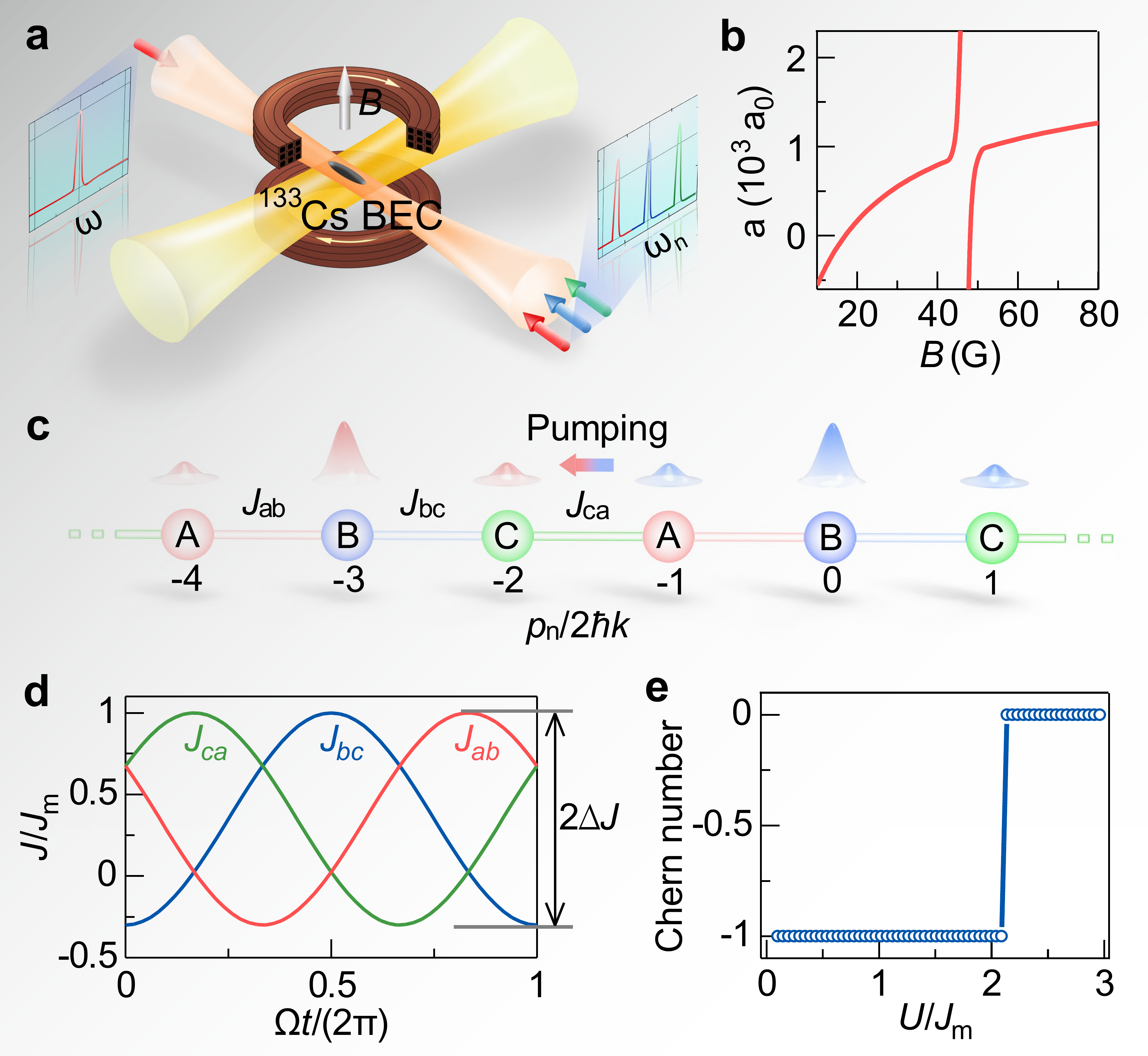}
\caption{\label{Fig1} Realization of Thouless pumping with tunable interaction. \textbf{a}, Sketch of experimental setup for coupling discrete momentum states of condensed $^{133}$Cs atoms via multiple counter-propagating Bragg laser pairs, whose frequency differences are controlled by engineering multi-frequency components $\omega_{n}$. \textbf{b}, Atomic interaction is adjusted by tuning the $s$-wave scattering length $a$, which varies with the magnetic field $B$ through a broad Feshbach resonance. \textbf{c}, Atomic discrete momentum states $p_{n}$ = 2$n\hbar k$ ($n = -9,\cdots,11$) are coupled by the Bragg laser pairs to realize an off-diagonal Aubry-Andr\'{e}-Harper model with three sites ($A$, $B$, $C$) per unit cell. \textbf{d}, Periodic modulation of coupling strengths during one pump cycle. \textbf{e}, Interaction-induced topological phase transition characterized by Chern numbers of Thouless pumping.}
\end{figure}

Here, we report the observation of interaction-induced fast quantized Thouless pumping of matter-wave solitons beyond conventional adiabatic limit. We experimentally study the synergistic effect between interaction and driving frequency in Thouless pumping, implemented in a periodically modulated momentum-space lattice of an atomic Bose-Einstein condensate (BEC). We show that increasing interaction strength significantly enlarges the frequency window supporting quantized pumping. In particular, the speed of quantized pumping is dramatically enhanced under intermediate interactions, which is absent in non-interacting or weakly interacting regimes. Upon further increasing the interaction strength, the system undergoes a topological phase transition, beyond which solitons become trapped and adiabatic pumping is suppressed. Remarkably, within a finite interaction range beyond the phase boundary, quantized soliton pumping reemerges at high modulation frequencies, while remaining absent in the low-frequency regimes. Our findings reveal a counterintuitive nonadiabatic yet quantized Thouless pumping mechanism.

\begin{figure*}
\includegraphics[width=17.5cm]{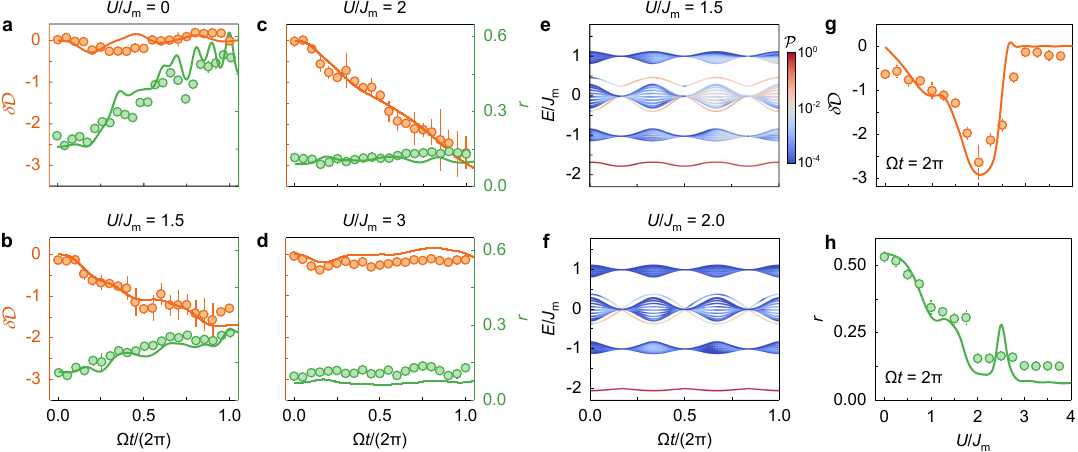}
\caption{\label{Fig2} Interaction-driven fast quantized Thouless pumping. \textbf{a}-\textbf{d}, Measured atomic center-of-mass displacement $\delta\mathcal{D}$ (orange circles) and participation ratio $r$ (green circles) as a function of evolution time $\Omega t$ for different interactions $U/J_{m}$.
\textbf{e},\textbf{f}, Numerically calculated instantaneous energy bands for $U/J_{m}$ = 1.5 and 2 under the mean-field approximation and periodic boundary condition. Color scale represents the projection of time-evolved state $|\phi(t)\rangle$ onto the instantaneous eigenstates $|\psi_{\rm NL}(t)\rangle$. \textbf{g},\textbf{h}, Variations of $\delta\mathcal{D}$ and $r$ measured after a pump cycle $\Omega t$ = 2$\pi$ with $U/J_{m}$. The solid lines in \textbf{a}-\textbf{d}, \textbf{g} and \textbf{h} are numerical simulations based on Eq.~(\ref{eq:Hint}). The error bars denote standard deviations. The modulation frequency and amplitude are $\Omega/J_{m}$ = 0.5 and $\Delta J/J_{m}$ = 0.65 with the maximum coupling strength $J_{m}/\hbar$ = $2\pi \times 1$kHz.}
\end{figure*}

\bigskip
\noindent \textbf{Aubry-Andr\'{e}-Harper model with tunable interactions} --- We experimentally realize Thouless pumping by implementing an Aubry-Andr\'{e}-Harper (AAH) model~\cite{Aubry:AIPS1980, Harper:PPSA1955} in a periodically modulated momentum-state lattice of $^{133}$Cs BEC with tunable interactions (see Figs.~\ref{Fig1}a and \ref{Fig1}b)~\cite{Weber:Sci2003}. Multiple Bragg laser pairs with wavelength $\lambda = 1064$nm are used to couple the 21 discrete momentum states $p_{n}= 2n\hbar k$ ($n \in \mathbb{Z}$) to form a 1D atomic chain~\cite{Meier:PRA2016, Wang:PRL2022}, where $\hbar$ is the reduced Planck's constant and $k = 2\pi/\lambda$ is the wave vector. The coupling strength is individually addressed by controlling the amplitude of corresponding Bragg laser field (see Methods). We then implement an off-diagonal AAH model with three sites ($A$, $B$ and $C$) per unit cell and zero on-site detuning~\cite{Jurgensen:Nat2021} by periodically modulating the coupling strengths as $J_{n} = J_{0} + \Delta J \cos\left(2n\pi/3 + \Omega t + \pi/3\right)$ (see Figs.~\ref{Fig1}c and \ref{Fig1}d), where $\Omega$ is the modulation frequency and  $J_{m}$ = $J_{0}$ + $\Delta J$ is the maximum coupling strength. The band structure of the 1+1D AAH model implemented in our experiment consists of three bands characterized by Chern numbers of $\mathcal{C} = \{-1, 2, -1\}$.

Originating from short-range interatomic interactions in real space, the interactions become long-ranged in the momentum space~\cite{An:PRL2018, Xie:PRL2020, An:PRL2021, Wang:PRL2022, Li:NC2023}. The impact of interactions on the Thouless pumping can be captured by the discrete nonlinear Schr\"{o}dinger equation
\begin{eqnarray}\label{eq:Hint}
		i\frac{\partial}{\partial t}\phi_{n}(t) &=& \sum_mH^{nm}_{\rm NL}[P_n(t),t]\phi_{m}(t).
\end{eqnarray}
Here, $\phi_{n}$ is the atomic wavefunction at lattice site $n$ and $H^{nm}_{\rm NL}$ is a nonlinear Hamiltonian defined as $H_{\rm NL}^{nm}=H_0^{nm}-UP_n(t)\delta_{m,n}$, where $H_0^{nm}=J_n\delta_{m,n+1}+J_{n-1}\delta_{m,n-1}$ is the AAH Hamiltonian, $P_n(t)=|\phi_n(t)|^2$ is the normalized atomic population at site $n$, and $U$ is the mean-field interaction energy that is proportional to the atomic $s$-wave scattering length $a$.
The atomic interactions for $a>0$ give rise to a density-dependent and local-attractive potential, leading to the nonlinear self-trapping soliton when $U$ is strong enough~\cite{Raghavan:PRA1999, Albiez:PRL2005}. Different from the focusing Kerr nonlinearity in optical systems~\cite{Jurgensen:Nat2021, Jurgensen:NP2023}, the nonlinearity directly results from the atomic collisional interactions, which are adjusted by tuning $a$ via a broad Feshbach resonance of $^{133}$Cs atoms~\cite{Chin:RMP2010}, as shown in Fig~\ref{Fig1}b. By combining the AAH model with tunable interactions, we realize a nonlinear Thouless pumping model in ultracold atoms. Within mean-field approximation, we evaluate the Chern number of the lowest band in nonlinear Thouless pumping~\cite{Niu:JPA1984} as a function interaction in Fig.~\ref{Fig1}e.

\bigskip
\noindent \textbf{Fast quantized Thouless pumping} --- Unlike linear Thouless pumping, quantized pumping of solitons has been recently observed in nonlinear photonic Thouless pumps~\cite{Jurgensen:Nat2021}. Solitons are pumped by an integer number of unit cells after an adiabatic pump cycle, as dictated by the Chern number of populated band. In our experiment, an exponentially localized Wannier state $|\phi(0)\rangle$ of the lowest band of Hamiltonian $H_{\rm NL}(k)$ is prepared by controlling atomic population distribution and their relative phases to initialize the system (see Methods and Supplementary Information).
We then quench the initialized state to the AAH model under various interactions, and measure the topological bulk dynamics of atoms in periodically modulated lattices. To quantitatively characterize the interaction effects on Thouless pumping, we define the center-of-mass (COM) displacement relative to the initial state as
\begin{eqnarray}\label{eq:COM}
		\delta\mathcal{D} &=& \sum_{n} n \times \left[P_{n}(t)-P_n(0)\right].
\end{eqnarray}

In the weakly interacting regime, $|\phi(0)\rangle$ corresponds to a mixed state with an approximately uniform population of the lowest band of $H_0$ (see Supplementary Information).
There is no unique instantaneous eigenstate that the system can follow~\cite{Song:SA2024}, and thus no direct criterion to define adiabatic pumping. Instead, the degree of adiabaticity is typically inferred from the deviation from quantized transport~\cite{Privitera:PRL2018}.
In contrast, in the regime of strong interactions, $|\phi(0)\rangle$ coincides with the lowest eigenstate of nonlinear Hamiltonian (see Figs.~\ref{Fig2}e and \ref{Fig2}f).
In this case, the adiabaticity can be directly quantified by evaluating the overlap between the time-evolved states and the instantaneous nonlinear eigenstate.

To characterize the interaction-induced localization associated with soliton formation, we introduce the participation ratio
\begin{eqnarray}\label{eq:Pr}
		r &=&\frac{1}{L}\frac{1}{\sum_{n}P_{n}^{2}},
\end{eqnarray}
where $L$ is the lattice size. For a diffusive wavepacket, $r$ increases from an intially small value during time evolution, and eventually approaches its maximum value of 1.
In contrast, for a soliton state, $r$ remains close to its minimum value throughout the pump cycle.

In Figs.~\ref{Fig2}a-\ref{Fig2}d, we show the measured dynamical evolution (points) and numerical simulations (lines) of the atomic COM displacement $\delta\mathcal{D}$ and participation ratio $r$ over a pump cycle for different interactions $U/J_{m}$ at a high modulation frequency $\Omega/J_{m}$ = 0.5.
All parameters are given in units of maximum coupling strength $J_{m}/\hbar$ = $2\pi$ $\times$ 1kHz. In the non-interacting limit, the increase of $r$ with time indicates a diffusive spreading of atomic wavepacket, while the atomic COM displacement remains nearly constant with $\delta\mathcal{D}\thicksim0$ (Fig.~\ref{Fig2}a). The breakdown of quantized pumping is attributed to strong nonadiabatic effects caused by the high modulation frequency.
With increasing $U$, $\delta\mathcal{D}$ approaches the quantized value $\delta\mathcal{D}=-3$ after one pump cycle, and the diffusive expansion is significantly suppressed, as evidenced by a slower growth of $r$ (Fig.~\ref{Fig2}b) compared with the non-interaction case. For $U/J_{m}$ = 2, the well-localized solitons form with small $r$, and their quantized motions are clearly observed during a fast pump cycle (Fig.~\ref{Fig2}c).
For stronger interaction $U/J_{m}=3$, the solitons become trapped, characterized by $\delta\mathcal{D} \thicksim 0$ and $r \rightarrow 1/L$, indicating localization of all atoms at their initialized sites (Fig.~\ref{Fig2}d). Our observation of interaction-induced fast quantized pumping is consistent with the interaction-enhanced topological pumping indicated in recent works with Rydberg atoms~\cite{Huang:arXiv2025} and optical waveguides~\cite{Chaudhari:arXiv2025}.

\begin{figure*}
\includegraphics[width=15cm]{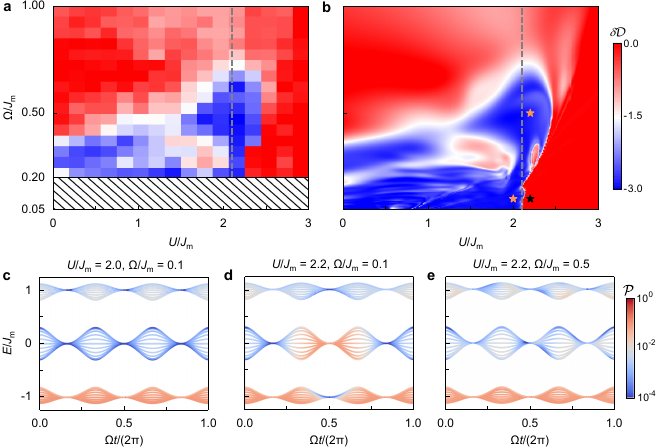}
\caption{\label{Fig3} Interaction-induced nonadiabatic quantized Thouless pumping. \textbf{a},\textbf{b}, Phase diagrams of Thouless pumping under the interplay between interaction $U/J_{m}$ and modulation frequency $\Omega/J_{m}$. Measured (\textbf{a}) and calculated (\textbf{b}) atomic center-of-mass displacement $\delta D$ obtained after a pump cycle as a function of $U/J_{m}$ and $\Omega/J_{m}$, where the hatched regime in \textbf{a} denotes the experimentally inaccessible low-frequency region due to the limitation of decoherence on the long-time propagation. Star points in \textbf{b} mark the parameters of $(U/J_m,\Omega/J_m)=(2,0.1)$, $(2.2,0.1)$ and $(2.2,0.5)$, respectively. Nonadiabatic quantized Thouless pumping emerges at the blue region beyond the phase boundary (dashed line). \textbf{c-e}, The projection of time-evolution state $|\phi(t)\rangle$ corresponding to the star points in \textbf{b} onto the linear eigenstates. The modulation amplitude is $\Delta J/J_{m}$ = 0.65 with $J_{m}/\hbar$ = $2\pi \times 1$kHz.}
\end{figure*}

To characterize the adiabaticity, we numerically calculate the instantaneous energy bands for $U/J_{m}=1.5$ and $2$ in Figs.~\ref{Fig2}e and \ref{Fig2}f, respectively. The color scale represents the projection $\mathcal{P}=|\langle\psi_{\rm NL}(t)|\phi(t)\rangle|^2$, where $|\psi_{\rm NL}(t)\rangle$ is the eigenstate of $H_{\rm NL}(t)$ and $|\phi(t)\rangle$ is the time-evolved state. In both cases, we identify low-energy soliton (lowest band) states that are separated from the linear bands by a finite gap, and $|\phi(t)\rangle$ coincides with the soliton state at $t=0$ due to the interactions. However, the subsequent evolution of $|\phi(t)\rangle$ exhibits distinct behaviors for different interactions.
For $U/J_{m}=1.5$, the projection $\mathcal{P}$ spreads over multiple bands during one pump cycle, indicating substantial nonadiabatic interband transitions that prevent quantized pumping. Consequently, only about $60\%$ of the population remains in the soliton band after one pump cycle. In comparison, the gap between soliton band and the higher bands increases with interaction, for $U/J_m=2$, the projection $\mathcal{P}$ remains predominantly within the soliton band throughout the pump cycle, reaching approximately $95\%$ at the end of one pump cycle and thereby enabling quantized transport. The measured (points) and calculated (lines) atomic COM displacement $\delta\mathcal{D}$ and participation ratio $r$ after a pump cycle as functions of $U/J_{m}$ are shown in Figs.~\ref{Fig2}g and \ref{Fig2}h, respectively. As discussed above, increasing $U/J_{m}$ drives the system from non-quantized pumping into quantized soliton pumping, before eventually entering a trapped phase under strong interactions.

\bigskip
\noindent \textbf{Nonadiabatic quantized Thouless pumping} --- For a more complete understanding of the interplay between interaction and modulation frequency in Thouless pumping, we systematically measure and simulate the atomic COM displacement $\delta\mathcal{D}$ after one pump cycle as a function of $U/J_{m}$ and $\Omega/J_{m}$, as shown in Figs.~\ref{Fig3}a and \ref{Fig3}b, respectively.
The hatched low-frequency region in Fig.~\ref{Fig3}a denotes the experimentally inaccessible regime due to the eventual decoherence during the long pumping period from the spatial separation of atoms excited to non-zero momentum states.
The dashed line indicates the boundary of interaction-induced topological phase transition at $U/J_{m}\sim 2.1$, characterized by the variation of Chern number with $U/J_{m}$ in Fig.~\ref{Fig1}e.
The phase diagram revealed by our experimental data is better verified by the numerical simulations, exhibiting three distinct regimes: (i) In the non- and weakly-interacting regimes, quantized pumping breaks down as $\Omega/J_{m}$ increases, due to strong nonadiabatic effects; (ii) With increasing $U/J_{m}$, the frequency window supporting quantized pumping is significantly enlarged, and the quantized transport persists even for high $\Omega/J_{m}$; and (iii) Beyond phase boundary (right side of dashed line), the solitons become trapped, leading to the suppression of quantized pumping in the low-frequency regime. However, in a broad region right beyond the phase boundary, we observe quantized pumping of solitons for high $\Omega/J_{m}$, while the solitons remain trapped for low $\Omega/J_{m}$. This reveals a nonadiabatic quantized pumping regime in the post-transition region, where the interactions enable the reemergence of quantized pumping at high frequencies.

\begin{figure*}
\includegraphics[width=16cm]{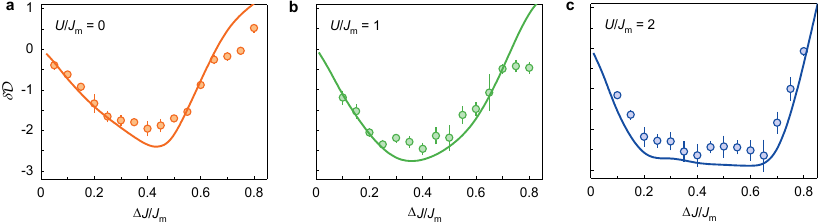}
\caption{\label{Fig4} Dependence of Thouless pumping on the modulation amplitude. \textbf{a-c}, The atomic center-of-mass displacement $\delta\mathcal{D}$ measured after a pump cycle $\Omega t$ = 2$\pi$ as a function of modulation amplitude $\Delta J/J_{m}$ for different interactions $U/J_{m}$. The solid lines are numerical simulations based on Eq.~(\ref{eq:Hint}). The error bars denote standard deviations. The modulation frequency is $\Omega/J_{m}$ = 0.5 with $J_{m}/\hbar$ = $2\pi \times 1$kHz.}
\end{figure*}

To elucidate the physical mechanism underlying this nonadiabatic quantized pumping, we analyze the overlaps of time-evolved state $|\phi(t)\rangle$ corresponding to the three star points in Fig.~\ref{Fig3}b, with the eigenstates of linear Hamiltonian $H_0$ in Figs.~\ref{Fig3}c-\ref{Fig3}e (corresponding adiabatic criterion can be found in Supplemental Information).
As shown in Fig.~\ref{Fig3}c, $|\phi(t)\rangle$ remains predominantly in the lowest band with Chern number $\mathcal{C}=-1$ throughout an adiabatic pump cycle for $U/J_{m}$ = 2, which explains the quantized pumping of solitons at the adiabatic (blue) region before the phase boundary in Fig.~\ref{Fig3}b.
By contrast, for $U/J_m=2.2$, $|\phi(t)\rangle$ initially populated at the lowest band is transferred to the second band with the Chern number $\mathcal{C}=2$ in Fig.~\ref{Fig3}d, and returns to the lowest band at the end of pump cycle. This interband transition causes the soliton trapping and breakdown of quantized pumping in the adiabatic (red) regime beyond the phase boundary.

Remarkably, as demonstrated in Fig.~\ref{Fig3}e, the nonadiabatic evolution at a high modulation frequency $\Omega/J_{m}$ = 0.5 effectively suppresses the interaction-induced interband transfer for $U/J_{m}$ = 2.2, i.e., $|\phi(t)\rangle$ is confined to the eigenstates of the lowest band during the entire nonadiabatic pump cycle. This synergetic interplay between nonadiabaticity and interaction stabilizes an effective single-band evolution, and enables nonadiabatic quantized pumping of solitons at the nonadiabatic (blue) region beyond the phase boundary in Fig.~\ref{Fig3}b. The breakdown of quantized pumping at a region centered at $(U/J_{m}, \Omega/J_{m})$ $\thicksim$ (1.75, 0.35) in Fig.~\ref{Fig3}b is attributed to the population on multiple bands during the pump cycle (see Supplementary Information).

\bigskip
\noindent \textbf{Interaction-relaxed quantized pumping condition} --- Besides, the modulation amplitude plays a crucial role in Thouless pumping. In Figs.~\ref{Fig4}a-\ref{Fig4}c, we plot the displacement $\delta\mathcal{D}$ measured after a pump cycle at $\Omega/J_{m}$ = 0.5 as a function of the modulation amplitude $\Delta J/J_{m}$ for different $U/J_{m}$. In the non-interacting limit, although $\delta\mathcal{D}$ varies with $\Delta J/J_{m}$, the minimal $\delta\mathcal{D}$ is larger than the negative integer displacement $\delta\mathcal{D}=-3$ due to the nonadiabatic effect. For $U/J_{m}$ = 1, $\delta\mathcal{D}$ decreases over a large range of $\Delta J/J_{m}$, and the minimal $\delta\mathcal{D}$ approaches the negative integer displacement of quantized pumping. For $U/J_{m}$ = 2, we observe the quantized pumping at most $\Delta J/J_{m}$, and the intermediate interaction relaxes the requirement of modulation amplitude to realize quantized Thouless pumping at high modulation frequency.\\

\noindent \textbf{Conclusion} ---
We investigate the interplay between interaction and modulation frequency in Thouless pumping realized in a periodically modulated lattice of an atomic BEC. We observe interaction-induced fast quantized  Thouless pumping of solitons, which is much more robust to nonadiabatic effect resulting from the finite pump period than the transport in the non- or weakly-interacting regimes. In particular, beyond the boundary of interaction-induced topological phase transition, we find an intriguing phenomenon of quantized pumping of solitons at high modulation frequencies, whereas solitons remain trapped in the low frequency regimes. This reveals a counterintuitive nonadiabatic yet quantized pumping of solitons, which is beyond the conventional Thouless pumping that requires the slow cyclic evolution of system's parameters to satisfy the adiabatic condition. Our findings offer new insights for understanding interaction effect on topological transport in periodically driving systems and may provide a route toward fast qubit control based on temporal modulation~\cite{Zhu:PRX2025}. Our work demonstrates a new strategy for developing compact topological devices by relaxing the adiabatic constraint in topological pumps in photonic waveguides~\cite{Jurgensen:Nat2021, Jurgensen:NP2023, Song:SA2024}.\\

\bigskip
\noindent \textbf{Methods}\\
\noindent \textbf{Experimental implementation}\\
We prepare a BEC of $^{133}$Cs atoms polarized in the hyperfine state $|F=3, m_{F}=3\rangle$ in a cigar-shaped optical trap~\cite{Wang:PRL2022}, and the trapping frequencies are $(\omega_{x}, \omega_{y}, \omega_{z})$ = 2$\pi$ $\times$ (125, 96, 13)Hz. To create the momentum-state lattice, the laser beam (wavelength $\lambda$ = 1064 nm) that provides the strongly radial confinement of optical trap is retro-reflected to form a pair of counter-propagating lasers for illuminating the weakly trapped BEC. We use two acousto-optic modulators to imprint the multiple-frequency components $\omega_{n}$ on the retro-reflected laser beam, while the incident laser beam has a single frequency $\omega$ (see Fig.~\ref{Fig1}a). These multiple-frequency Bragg laser pairs, far detuned from the atomic transition, couple the discrete momentum states with the increment of $2\hbar k$~\cite{Meier:PRA2016}. According to the quadratic energy-momentum dispersion, the Bragg transition frequency between the nearest-neighbor momentum states $p_{n}$ and $p_{n+1}$ is given by
\begin{eqnarray}\label{eq:EP}
		\Delta \omega = \omega_{n}-\omega = (2n+1)4E_{R}/\hbar,
\end{eqnarray}
where $E_{R}$=$\hbar^{2}k^{2}/2m$ is the one-photon recoil energy with the atomic mass $m$.

Considering the total power of multiple Bragg laser pairs, we can synthesize a lattice with the size $L$ $>$ 50 sites, but the condensed atoms cannot populate all of these sites within the available evolution duration, due to the eventual decoherence from the spatial separation of atoms in different momentum states. In this work, we drive 20 Bragg transitions, and synthesize a 21-site lattice based on the laser-coupled momentum states.  Since each Bragg transition has a unique resonant frequency, we can control the amplitudes and phases of all transitions as well as the detunings from Bragg resonances by individually addressing the corresponding frequency components, and thus realize the site-resolved control of the coupling strength, tunneling phase and on-site energy for the synthetic lattice. When  the coupling strengths are periodically modulated with a trimer spatial structure (see Fig.~\ref{Fig1}d), we implement an AAH model to realize quantized Thouless pumping.\\

\noindent \textbf{Discrete nonlinear Schr\"{o}dinger equation}\\
Different from the short-range interatomic interactions in the real space, the atomic interactions become long-ranged in the momentum space.
Under the mean-field approximation, the impact of atomic interactions on the Thouless pump is described by the discrete nonlinear Schr\"{o}dinger equation~\cite{An:PRL2018, Xie:PRL2020, An:PRL2021, Wang:PRL2022, Li:NC2023}
\begin{eqnarray}
\label{eq:GPE1}
i\hbar\frac{\partial\phi_n}{\partial t}&=-& J_{n}\phi_{n+1} - J_{n-1}\phi_{n-1} \nonumber\\
&+&U\Big(|\phi_n|^2+\sum_{\ell \neq n}2|\phi_{\ell}|^2\Big)\phi_{n}.
\end{eqnarray}
Here we restrict the interaction processes to mode-conserving collisions, and retain only density-density terms~\cite{An:PRL2021}, including both the on-site and inter-site contributions.
We further neglect the interaction processes in which atoms are scattered out of the discrete momentum modes resonantly coupled by the Bragg laser pairs, as well as scattering process involving more than two sites, which are energetically suppressed.
Due to the bosonic exchange statistics, a pair of atoms occupying different sites acquires an additional exchange energy compared to the case where both atoms occupy on the same site.
As a result, the inter-site interaction is twice as strong as the on-site interaction, as described in the interaction terms in Eq.~(\ref{eq:GPE1}). This gives rise to an effective on-site attractive interaction in momentum-state lattices. Considering the normalization condition $\sum_{n}|\phi_n|^2 = 1$, Eq.~(\ref{eq:GPE1}) can be simplified to obtain the nonlinear Schr\"{o}dinger equation (see Eq.~(\ref{eq:Hint})) after neglecting an irrelevant overall energy shift.\\

\noindent \textbf{Preparation of initial Wannier state}\\
To realize quantized Thouless pumping in the AAH model for various interactions, we prepare the initial state $|\phi(0)\rangle$ of the system as the maximally localized Wannier state of the lowest band of $H_{\rm NL}^{ab}(k,0)=H_0(k)-\delta_{ab}UP_a(k)$, where $H_0$ is the linear AAH Hamiltonian in momentum space, $a,b \in \{A,B,C\}$ denote the sublattice indices in each unit cell, and $P_a(k)$ denotes the self-consistent sublattice population obtained by iteratively solving the $H_{\rm NL}$ starting from the lowest band Bloch eigenstate of $H_0(k)$.
This initial state is exponentially localized in lattice space, while uniformly populating the lowest band of $H_{\rm NL}(k,0)$.
For the lowest band with Bloch eigenstate $|u_{1}(k)\rangle$, the phase is chosen to ensure the smooth variation of $|u_{1}(k)\rangle$ with $k$.
The corresponding Wannier function can then be expressed as $|\phi_j(0)\rangle = \int_{-\pi}^{\pi}e^{ikj}u_{1}(k)\,{\rm d}k/2\pi$ (see Supplementary Information).
We find that the initial Wannier state mainly populates over three neighboring sites with a $\pi$-phase difference between the center and two side sites.
Moreover, as the interaction increases, the population at the center site increases while that at the side sites decreases.

In the experiment, we start with all couplings turned off, and all atoms are prepared at the center zero-momentum site. Taking advantages of the site-resolved control over the coupling strengths and tunneling phases in momentum-state lattices, we engineer the initial population distribution of atoms in the lattice to prepare the Wannier state of the lowest band for different interactions. We use two Bragg pulses to create the initial population at the three neighboring sites with the controlled proportion and phase, and this initialization is approximate to the population distribution determined by the calculated Wannier function. Then we study the topological bulk dynamics after quenching the initial Wannier state to the AAH model under different interactions.\\

%\noindent \textbf{Experimental detection}\\
%For detection, all laser fields are switched off after an evolution time, and the magnetic field is tuned to the zero-crossing point of the scattering length with $B$ = 17 G. The atomic distribution is then measured through the standard absorption imaging after a 22 ms time-of-flight, during which the atoms with distinct momenta naturally separate in space. Through Gaussian fitting, we select the distribution region of the atoms populating at each momentum state, and then extract the population of atoms at the corresponding lattice site.\\

\noindent \textbf{Acknowledgement}

\noindent We acknowledge Yongguan Ke, Chaohong Lee, Sanyi You and Yedi Shen for helpful discussions on the manuscript. This work is supported by the Innovation Program for Quantum Science and Technology (Grant Nos. 2021ZD0302103 (J.M.), 2021ZD0302800(Z.Q.)), the National Key Research and Development Program of China (Grant No. 2022YFA1404201 (L.X.)),  the National Natural Science Foundation of China (Grant Nos. 62325505 (J.M.), 62422508 (Y.L.), 62175140 (Y.W.), 12488101(Z.Q.)). We also thank the Supercomputing Center of University of Science and Technology of China for providing high-performance computing resources.\\

\noindent\textbf{Author contributions}

\noindent Y.L., Y.W., H.D. and J.M. contributed to the executions of the experiments. J.J, Z.A., Z.Q. and Q.N. developed the theoretical model. L.X., S.J. and J.M. supervised the project. All authors discussed the results, contributed to the data analysis and co-wrote the manuscript.\\

\noindent \textbf{Competing interests}\\
The authors declare no competing interests.\\

\noindent \textbf{Data availability}\\
The data that support the findings of this study are available from the corresponding authors upon request. \\

\noindent \textbf{Code availability}\\
The codes that support the findings of this study are available from the corresponding authors.\\

\end{document}